\documentclass[preprint,12pt]{elsarticle}
\usepackage{amssymb}
\journal{J. Magn. Magn. Mater.}

\begin{document}

\begin{frontmatter}

\title{
Inelastic neutron scattering study on 
the AFM uniform spin-$\frac{1}{2}$ chain compound CuSb$_2$O$_6$ 
}

\author[inst1]{Masashi Hase}
\author[inst2]{Minoru Soda}
\author[inst3]{Takatsugu Masuda}
\author[inst4]{Shinichi Itoh}
\author[inst4,inst5,inst6,inst7]{Tetsuya Yokoo}

\address[inst1]{
Research Center for Materials Nanoarchitectonics (MANA), 
National Institute for Materials Science (NIMS),
Tsukuba, 305-0047, Ibaraki, Japan 
}
\address[inst2]{
Department of Physics, Advanced Sciences, G.S.H.S.,  
Ochanomizu University,
Bunkyo, 112-8610, Tokyo, Japan
}
\address[inst3]{
The Institute for Solid State Physics (ISSP), 
The University of Tokyo,
Kashiwa, 277-8581, Chiba, Japan
}
\address[inst4]{
Institute of Materials Structure Science (IMSS), 
High Energy Accelerator Research Organization (KEK),
Tsukuba, 305-0801, Ibaraki, Japan
}
\address[inst5]{
Neutron Science Section, Materials and Life Science Division, 
J-PARC Center, 
Tokai, 319-1195, Ibaraki, Japan
}
\address[inst6]{
Materials Structure Science, Department of Advanced Studies, 
The Graduate University for Advanced Studies SOKENDAI,
Tsukuba, 305-0801, Ibaraki, Japan
}
\address[inst7]{
Degree Programs in Pure and Applied Sciences, Graduate School of Science and Technology,
 University of Tsukuba,
Tsukuba, 305-8573, Ibaraki, Japan
}


\begin{abstract}

We carried out inelastic neutron scattering experiments on a powdered sample of 
the antiferromagnetic (AFM) uniform spin-$\frac{1}{2}$ chain compound 
CuSb$_2$O$_6$. 
The magnetic excitations appear in the energy range  
$1.8 \le \omega \le 13$~meV at 2.5 K 
below the AFM transition temperature ($T_{\rm N} = 8.7$~K). 
The gap value (1.8 meV) is close to that evaluated from the specific heat 
(1.51 meV).
The excitations at 12.5 K ($> T_{\rm N}$) appear gapless. 
Thus, 
the 1.8 meV gap is caused by some anisotropy in spin-wave excitations. 
The gap excitations are strongest at $Q_{\rm zc} \approx 0.48$ \ \AA$^{-1}$, 
corresponding to a Cu--Cu length of 6.6 \AA. 
This result is consistent with the theoretical one that 
the interaction in a Cu--Cu pair with a length of 6.5562 \AA \ ($J_{ab}$)
is strongest. 
The magnetic excitations can be explained by 
the AFM uniform XXZ chain with $J_{ab} = 6.437$~meV and 
$\Delta J_{ab} = 0.063$~meV. 
The 1.8 meV gap is caused by 
the small Ising anisotropy ($\frac{\Delta J_{ab}}{J_{ab}} = 0.0098$).

\end{abstract}



\end{frontmatter}

\section{Introduction}

Precise knowledge of exchange interactions is clearly a prerequisite 
for understanding the magnetic properties of quantum spin systems. 
Strong exchange interactions usually exist in short {\it M}--{\it M} pairs, 
where {\it M} denotes a magnetic ion such as a Cu$^{2+}$ ion. 
According to the Kanamori--Goodenough 
rule \cite{Kanamori57a,Kanamori57b,Kanamori59,Goodenough55,Goodenough58,Anderson50}, 
an antiferromagnetic (AFM) superexchange interaction is stronger 
when its {\it M}--O--{\it M} angle is closer to 180$^{\circ}$. 
For a given crystal structure, 
we infer which {\it M}--{\it M} pairs can have strong exchange interactions.   
In some materials, however, 
a long {\it M}--{\it M} pair has 
an interaction on the order of several tens of kelvin. 
The interaction is possible in 
an {\it M}--O--O--{\it M} or {\it M}--O--{\it A}--O--{\it M} 
({\it A} = nonmagnetic ion) path.  
Examples of such materials include 
CaCuGe$_2$O$_6$ \cite{Sasago95,Zheludev96}, CuWO$_4$ \cite{Lake96}, 
VODPO${_4}\cdot$1/2D$_2$O \cite{Tennant97}, (VO)$_2$P$_2$O$_7$ \cite{Garrett97}, 
and $\beta$-AgCuPO$_4$ \cite{Hase07}. 
Some theoretical results have shown that 
a long {\it M}--{\it M} pair can have 
a large interaction \cite{Valenti02,Yahia06}. 

We have been focusing on CuSb$_2$O$_6$. 
The strongest exchange interaction is 96 K (8.3 meV), whereas 
the shortest Cu--Cu bond length is 4.635 \AA.
Figures 1(a) and 1(b) depict the crystal structure of CuSb$_2$O$_6$. 
The space group is monoclinic $P2_1/n$, and the lattice constants are 
$a=4.6349(1)$~\AA, $b=4.6370(1)$~\AA, $c=9.2931(10)$~\AA, and 
$\beta = 91.124(2)^{\circ}$ \ at room temperature \cite{Nakua91,Ramos91}.
Figures 1(c)--(e) show a CuO$_6$ octahedron projected in three directions. 
The Cu--O lengths and O--Cu--O angles are described in the caption. 
The distortion of the octahedron from cubic symmetry
is small \cite{Nakua91,Ramos91,Giere97}. 
The small Jahn--Teller distortion indicates that 
the energies of the $3d_{3z^2-r^2}$ and $3d_{x^2-y^2}$ orbitals are similar to each other \cite{Kasinathan08}.
\begin{figure} \begin{center}
\includegraphics[width=12cm]{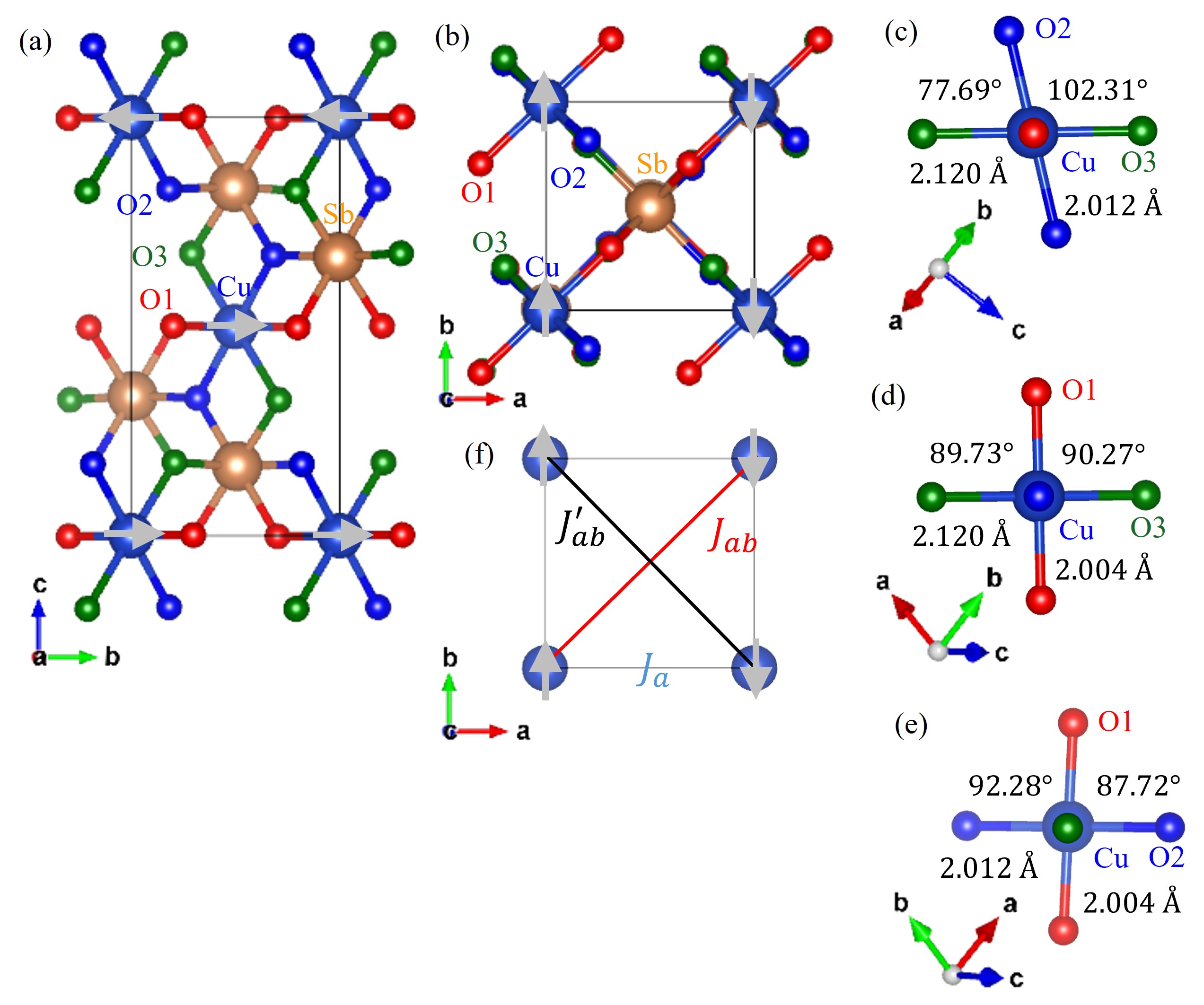}
\caption{
(a)--(b)
The crystal structure of CuSb$_2$O$_6$ \cite{Nakua91,Ramos91}, 
as drawn using VESTA \cite{Momma11}. 
The space group is monoclinic $P2_1/n$ (No. 14), and the lattice constants are 
$a=4.6349(1)$~\AA, $b=4.6370(1)$~\AA, $c=9.2931(10)$~\AA, and 
$\beta = 91.124(2)^{\circ}$ \ at room temperature. 
We use $P2_1/n$ in this paper because it is the most commonly used space group for CuSb$_2$O$_6$ in related papers.  In the standard notation,   
space group No. 14 is expressed as $P2_1/c$ with 
$a=4.6349(1)$~\AA, $b=4.6370(1)$~\AA, $c=10.3031(10)$~\AA, and 
$\beta = 115.605(2)^{\circ}$. 
There are one Cu$^{2+}$, one Sb$^{5+}$, and three O$^{2-}$ 
crystallographic sites.
The gray box represents the crystallographic unit cell. 
The arrows show the ordered magnetic moments \cite{Kato02}.  
(c)--(e)
A CuO$_6$ octahedron projected in three directions. 
Cu--O lengths and O--Cu--O angles are also described.
The O$i$--Cu--O$i$ \ ($1 \le i \le 3$) angles are 180$^{\circ}$.
(f)
Candidates for the exchange interaction forming the AFM uniform chain. The Cu--Cu length is 4.6349, 6.5562, and 6.5562 \AA \ for 
the $J_a$, $J_{ab}$, and $J'_{ab}$ interactions, respectively. 
}
\end{center}
\end{figure}

The AFM transition was observed at $T_{\rm N} = 8.7$~K 
in the temperature $T$ dependence of 
the magnetic susceptibility and specific heat \cite{Nakua91,Gibson04,Rebello13}.  
The magnetic susceptibility at $T > T_{\rm N}$ 
agrees well with that of the AFM uniform spin-$\frac{1}{2}$ chain 
with the exchange interaction value $J = 96$~K 
(8.3 meV) \cite{Nakua91,Kato02,Rebello13}.  
Here, the exchange interaction parameter $J$ is defined in the Hamiltonian 
$J \sum_i {\bf S}_i \cdot {\bf S}_{i+1}$. 
The magnetic susceptibility at $T > T_{\rm N}$ is nearly isotropic. 
The $g$ values parallel to the $a$, $b$, and $c$ directions were evaluated to be 
$g_a = 2.20(1)$, $g_b = 2.20(1)$, and $g_c = 2.24(2)$, 
respectively \cite{Rebello13}. 
These $g$ values are close to those obtained in electron-spin resonance (ESR) measurements at 20 K 
($g_a = 2.21$, $g_b = 2.21$, and $g_c = 2.23$) \cite{Heinrich03}. 
As $T$ is lowered below $T_{\rm N}$, 
the magnetic susceptibility parallel to the $a$, $b$, and $c$ directions
slightly decreases, rapidly decreases, and is nearly constant, 
respectively \cite{Gibson04,Rebello13,Herak15}. 
The spin-flop field is 1.25 T \cite{Rebello13,Heinrich03}. 

The magnetic structure is shown in Figs. 1(a) and 1(b).
The propagation vector is 
${\bf k} = (1/2, 0, 1/2)$ \cite{Kato02,Gibson04,Nakua95}. 
Kato {\it et al.} reported that the ordered magnetic moments are parallel to 
the $b$ direction \cite{Kato02}, whereas 
Gibson {\it et al.} reported that the main component of the moments is $b$ and 
that a small $a$ component exists \cite{Gibson04}. 
The direction of the ordered moments is consistent with 
the anisotropic magnetic susceptibility at 
$T < T_{\rm N}$ \cite{Gibson04,Rebello13,Herak15}. 
Given the magnetic structure, 
candidates of the exchange interaction forming the AFM uniform chain are 
the $J_a$, $J_{ab}$, and $J'_{ab}$ interactions shown in Fig. 1(f).
According to the theoretical results of Kasinathan {\it et al.} \cite{Kasinathan08} and Koo and Whangbo \cite{Koo01}, 
the AFM $J_{ab}$ interaction is strongest. 

The magnetic specific heat at $T < T_{\rm N}$ is reproduced by 
the formula $A \exp(-\Delta/T)$ with $\Delta = 17.5$~K (1.51 meV) 
at zero magnetic field, suggesting the existence of an energy gap 
in the magnetic excitations \cite{Rebello13}. 
The corresponding nuclear magnetic resonance (NMR) results show that the spin-lattice relaxation rate ($1/T_1$) 
at $T < T_{\rm N}$ also exhibits gap-like behavior \cite{Kato02}.
Both the specific heat and the linear thermal expansion coefficients
exhibit distinct jumps at $T_{\rm N}$ \cite{Rebello13}. 
The unusual character of this phase transition suggests that 
it is a crossover between one-dimensional antiferromagnetism and 
three-dimensional antiferromagnetic order 
possibly associated with a spin-Peierls 
transition \cite{Hase93a,Hase93b,Hase93c}. 

It is important to determine 
which exchange interaction forms the AFM uniform chain and 
to confirm whether the gap exists in the magnetic excitations. 
If the gap exists, it is necessary to investigate its origin. 
We should confirm whether or not the gap can be explained by some anisotropy or 
we should consider another origin such as the spin-Peierls transition.
Consequently, we carried out inelastic neutron scattering (INS) experiments on 
CuSb$_2$O$_6$ powder. 

\section{Experimental and Calculation Methods}

We synthesized crystalline powder of CuSb$_2$O$_6$ via a solid-state reaction.
The starting materials were 
CuO and Sb$_2$O$_5$ powders with 99.99\% purity.
A stoichiometric mixture of the powders was sintered at 1273 K 
in air with intermediate grinding. 
We recorded the X-ray powder diffraction patterns for the product at room temperature
using an X-ray diffractometer (RINT-TTR III; Rigaku) and 
confirmed the formation of the target compound. 

We carried out INS measurements using 
the High-Resolution Chopper (HRC) spectrometer at beamline BL 12 
in the Japan Proton Accelerator Research Complex 
(J-PARC) \cite{Itoh11,Yano11,Itoh12b}.
We calculated dispersion relations of spin-wave excitations using 
linear spin-wave theory realized in 
the SpinW program package \cite{Toth15}. 

\section{Results and Discussion}

Figure 2 shows the INS intensity $I(Q, \omega)$ map for CuSb$_2$O$_6$ powder, 
where $Q$ and $\omega$ are
the magnitude of the scattering vector and the energy transfer, respectively.
The energy of the incident neutrons ($E_{\rm i}$) is 
11.4 meV in Figs. 2(a) and 2(c) and 
51.0 meV in Figs. 2(b) and 2(d).
Excitations in the energy range 
$1.8 \le \omega \le 13$~meV are observed at 2.5 K [Figs. 2(a) and 2(b)]. 
The excitations at $\omega \approx 13$~meV are 
magnetic ones at the zone boundary of the AFM uniform chain 
because the value of $\frac{\pi}{2} J$ is 13 meV \cite{Cloizeau62}. 
No excitation is observed below 1.8 meV, 
indicating the existence of an excitation gap.
This gap value (1.8 meV) is close to that evaluated from the specific heat 
(1.51 meV) \cite{Rebello13}. 
As shown in Fig. 2(c), 
the excitations at 12.5 K ($> T_{\rm N} = 8.7$~K) appear to be gapless. 
We infer that 
the 1.8 meV gap is caused by some anisotropy in spin-wave excitations. 
The gap excitations are strongest at $Q_{\rm zc} \approx 0.48$ \ \AA$^{-1}$, 
corresponding to a Cu--Cu length of 6.6 \AA. 
This result is consistent with the theoretical one indicating that 
$J_{ab}$, for which the Cu--Cu length is 6.5562 \AA, is 
strongest \cite{Kasinathan08,Koo01}.
The excitations at 1.8 meV exist at $Q > Q_{\rm zc}$ 
because signals at $Q > Q_{\rm zc}$ can include the excitation at $Q_{\rm zc}$ 
in the powder sample. 


\begin{figure}\begin{center}
\includegraphics[width=12cm]{JMMM05_Fig2.jpg}
\caption{
The INS intensity $I(Q, \omega)$ maps of CuSb$_2$O$_6$ powder. 
The measurement temperature is 2.5 K for (a) and (b) and 12.5 K for (c) and (d).  
$E_{\rm i}$ \ is 11.4 meV for (a) and (c) and 51.0 meV for (b) and (d).  
The solid line in (a) and (b) shows the dispersion relation along the chain direction of 
the AFM uniform XXZ chain with 
$J_{ab} = 6.437$~meV and $\Delta J_{ab} = 0.063$~meV. 
The vertical key on the right shows the INS intensity in arbitrary units. 
}
\end{center}
\end{figure}

We calculated spin-wave excitations for the following AFM uniform XXZ chain 
using the SpinW program package \cite{Toth15}: 
\begin{equation}
{\cal H} = \sum_{i} (
J_{ab} {\bf S}_{i} \cdot {\bf S}_{i+1} + 
\Delta J_{ab} S_i^z S_{i+1}^z).
\end{equation}
We assumed that the interchain interactions are zero 
because 
the magnetic susceptibility at $T > T_{\rm N}$ 
can be explained well by the AFM uniform spin-$\frac{1}{2}$ 
chain \cite{Nakua91,Kato02,Rebello13}. 
The solid line in Figs. 2(a) and 2(b) shows 
the dispersion relation along the chain direction 
when $J_{ab} = 6.437$~meV and 
$\Delta J_{ab} = 0.063$~meV.  
The excitation energies at the zone center and boundary 
(1.8 and 13 meV, respectively) can be reproduced. 
The small Ising anisotropy ($\frac{\Delta J_{ab}}{J_{ab}} = 0.0098$) 
can explain the 1.8 meV gap. 
Notably, the zone boundary energy is $2(J_{ab} + \Delta J_{ab})$ 
in spin-wave excitations  
[not $\frac{\pi}{2} (J_{ab} + \Delta J_{ab})$ in AFM uniform spin-$\frac{1}{2}$ chains]. 
Wheeler has reported similar results \cite{Wheeler07}. 

We consider another anisotropy (single-ion anisotropy).
The spin-wave gap in Cu$_6$[Si$_6$O$_{18}$]$\cdot$6H$_2$O (1.5 meV)
can be explained by the Ising anisotropy of 
1.3\% \cite{Podlesnyak16}. 
However, the gap can also be explained on the basis of single-ion anisotropy. 
According to the theoretical results of Liu {\it et al.} \cite{Liu14}, 
single-ion anisotropy is generated by spin--orbit coupling 
even for $S=1/2$ spins. 
Spin--orbit coupling has been suggested to exist in CuSb$_2$O$_6$ \cite{Rebello13}. 
Thus, single-ion anisotropy in CuSb$_2$O$_6$ may be possible, 
However, as described in the Introduction, 
the anisotropy in real space is small. 

\section{Summary}

We performed INS experiments on a powdered sample of 
the AFM uniform spin-$\frac{1}{2}$ chain compound CuSb$_2$O$_6$. 
We observed the magnetic excitations in the energy range 
$1.8 \le \omega \le 13$~meV at 2.5 K ($< T_{\rm N} = 8.7$~K).
This gap value (1.8 meV) is close to that determined from the specific heat 
(1.51 meV).
The excitations at 12.5 K ($> T_{\rm N}$) appear to be gapless. 
Therefore, 
the 1.8 meV gap is generated by some anisotropy in spin-wave excitations. 
The gap excitations are strongest at $Q_{\rm zc} \approx 0.48$ \ \AA$^{-1}$, 
corresponding to a Cu--Cu length of 6.6 \AA. 
This result is consistent with the theoretical one showing that 
$J_{ab}$, for which the Cu--Cu length is 6.5562 \AA, is strongest. 
The excitation energy of 13 meV agrees well with 
the zone boundary energy of the AFM uniform chain ($\frac{\pi}{2} J$)
where the value of $J$ (8.3 meV) was determined from the magnetic susceptibility.  
The magnetic excitations can be explained by 
the AFM uniform XXZ chain with $J_{ab} = 6.437$~meV and 
$\Delta J_{ab} = 0.063$~meV. 
The 1.8 meV gap is caused by 
the small Ising anisotropy ($\frac{\Delta J_{ab}}{J_{ab}} = 0.0098$).

\section*{Acknowledgments}

This work was supported 
by the Japan Society for the Promotion of Science (JSPS) KAKENHI 
Grant Number 24K06947 and  
by the World Premier International Research Center Initiative (WPI), 
Ministry of Education, Culture, Sports, Science and Technology (MEXT), Japan.
The neutron scattering experiments were approved by the Neutron Science Proposal Review Committee of J-PARC/MLF (Proposal Number 2014B0023) and 
supported by the Inter-University Research Program on Neutron Scattering of IMSS, KEK.
We thank Seiko Matsumoto 
at the National Institute for Materials Science (NIMS) 
for the sample synthesis and X-ray diffraction measurements. 
We thank Daichi Kawana, Ryosuke Sugiura, and Toshio Asami
at the Neutron Science Laboratory, Institute for Solid State Physics (ISSP), The University of Tokyo 
for supporting the INS experiments. 
We are grateful to 
Hiroaki Mamiya, Masamichi Nishino, Kazunari Yamaura, Alexei Belik, and Yoshihiro Tsujimoto 
at NIMS for fruitful discussions. 





\end{document}